\documentclass[final,twocolumn]{IEEEtran}

\usepackage[cmex10]{amsmath}

\usepackage{latexsym}
\usepackage{graphics}
\usepackage{amssymb}
\usepackage{amsthm}

\usepackage{cite}
\usepackage{array}
\usepackage{graphicx}
\usepackage[linesnumbered,ruled,norelsize,vlined]{algorithm2e}

\usepackage{url}
\usepackage{color}
\usepackage{subfigure}
\usepackage{epstopdf}

\def\r{\mathbf{r}}
\def\a{\mathbf{a}}
\def\x{\mathbf{x}}
\def\R{\mathbf{R}}

\def\th{\boldsymbol{\theta}}
\def\atil{\mathbf{\tilde{a}}}
\def\ahat{\mathbf{\hat{a}}}

\newcommand{\mbf}[1]{\mathbf{#1}} 

\usepackage[cmex10]{amsmath}
\usepackage{cite}
\usepackage{array}
\usepackage{graphicx}
\usepackage[linesnumbered,ruled,norelsize,vlined]{algorithm2e}

\usepackage{url}
\usepackage{color}
\usepackage{subfigure}

\newtheorem{theo}{Theorem}
\newtheorem{lem}{Lemma}

\title{Robust Optimal Power Distribution for Hyperthermia Cancer Treatment}%
\author{Nafiseh Shariati, Dave Zachariah, Johan Karlsson, Mats Bengtsson}%

\begin{document}

\maketitle

\begin{abstract}
We consider an optimization problem for spatial power distribution generated by an array of transmitting elements. Using ultrasound hyperthermia cancer treatment as a motivating example, the signal design problem consists of optimizing the power distribution across the tumor and healthy tissue regions, respectively. The models used in the optimization problem are, however, invariably subject to errors. deposition as well as inefficient treatment. To combat such unknown model errors, we formulate a robust signal design framework that can take the uncertainty into account using a worst-case approach. This leads to a semi-infinite programming (SIP) robust design problem which we reformulate as a tractable convex problem, potentially has a wider range of applications.
\end{abstract}

\section{Introduction} \label{sec:introduction}

Local hyperthermia is a noninvasive technique for cancer treatment in which targeted body tissue is exposed to high temperatures to damage cancer cells while leaving surrounding tissue unharmed. This technique is used both to kill off cancer cells in tumors and as a means to enhance other treatments such as radiotherapy and chemotherapy. Hyperthermia has the potential to treat many types of cancer, including sarcoma, melanoma, and cancers of the head and neck, brain, lung, esophagus, breast, bladder, rectum, liver, appendix, cervix, etc. \cite{Falk&Inssels2001_hyperthermia,Zee2002_heating,WustEtAl2002_hyperthermia}.

Hyperthermia treatment planning involves modeling patient-specific tissue, using medical imaging techniques such as microwave, ultrasound, magnetic resonance or computed tomography, and calculating the spatial distribution of power deposited in the tissue to heat it \cite{13: Paulides}. There exist two major techniques to concentrate the power in a well-defined tumor region: electromagnetic and ultrasound, each with its own limitations. The drawback of electromagnetic microwaves is poor penetration in biological tissue, while for ultrasound the short acoustic wavelength renders the focal spot very small. Using signal design methods, however, one can improve the spatial power deposition generated by an array of acoustic transducers. Specifically, standard phased array techniques do not make use of combining a diversity of signals transmitted at each transducer. When this diversity exploited it is possible to dramatically improve the power distribution in the tumor tissue, thus improving the effectiveness of the method and reducing treatment time \cite{08:Guo,Jennings&McGough2010_improving}. Given a set of spatial coordinates that describe the tumor region and the healthy tissue, respectively, the transmitted waveforms can be designed to optimize the spatial power distribution while subject to certain design constraints.

One critical limitation, however, is the assumption of an ideal wave propagation model from the transducers to a given point in the tissue. Specifically, model mismatches may arise from hardware imperfections, tissue inhomogeneities, inaccurately specified propagation velocities, etc. Thus the actual power distribution may differ substantially from the ideal one designed by an assumed model. This results in suboptimal clinical outcome due to loss of power in the tumor region and safety issues due to the possible damage of healthy tissue. These considerations motivate developing robust design schemes that take such unknown errors into account.

In this paper we derive a robust optimization method that only assumes the unknown model errors to be bounded. The power is then optimized with respect to `worst-case' model errors. By using a worst-case model, we provide an optimal signal design scheme that takes into account all possible, bounded model errors. Such a conservative approach is warranted in signal design for medical applications due to safety and health considerations. Our method further generalizes the approach in \cite{08:Guo} by obviating the need to specify a fictitious tumor center point. The framework developed here has potential use in wider signal design applications where the resulting transmit power distributions are subject to model inaccuracies. More specifically, the design problem formulated in this paper and the proposed robust scheme can be exploited to robustify the spatial power distribution for applications that an array equipped with multiple elements is used to emit waveforms in order to deliver power to an area of interest in a controlled manner.

The core of this study is built upon exploiting waveform diversity which has been introduced in multiple-input multiple-output (MIMO) radar literature \cite{Stoica2007a}, and later has been applied for local hyperthermia cancer treatment improvement in \cite{08:Guo}. In the MIMO radar field, robustness studies have been carried out in different applications under varying design parameter uncertainties, cf., \cite{07:Yang,11:Grossi}. Recently, in \cite{14:Shariati2014c}, we have studied the robustification of the waveform diversity methodology for MIMO radar applications. It should be highlighted that in this paper a more generic problem formulation has been studied with respect to those of \cite{14:Shariati2014c}, where a new application area is considered to illustrate the performance of our proposed robust design. In the array processing literature, beamforming under array model errors has also spawned extensive work, cf., \cite{04:Li,08:Yan,08:Kim,12:Khabbazibasmenj}.

For hyperthermia therapy, the need for robust solutions when optimizing for phase and amplitude of conventional phased array, has been investigated in \cite{deGreef2010a} considering perfusion uncertainties, and in \cite{deGreef2011a} considering dielectric uncertainties. The authors emphasize on the role of uncertainty in such designs (hyperthermia planning) since it influences the calculation of power distribution, and correspondingly temperature distribution.

The paper is organized as follows: In Section~\ref{sec:system model}, we describe the system model and the relevant variables. In Section~\ref{sec:problem formulation}, the signal design problem is presented. First, we consider the state-of-the-art method based on `waveform diversity' \cite{08:Guo,Stoica2007a,JLi2009a}, then we generalize the design problem by introducing a deterministic and bounded set of possible model errors which results in an infinite number of constraints. Importantly we show that this seemingly intractable problem can be equivalently formulated as a tractable convex optimization problem. In Section~\ref{sec:numerical_results}, we evaluate  the design scheme. We evaluate the performance of our proposed robust power distribution scheme specifically for local hyperthermia breast cancer treatment. This example application is motivated by the alarming statistics pointing to breast cancer as one of the leading causes of death among women worldwide \cite{CancerReport2014UK,CancerReport2014US,CancerReport2013French}\footnote{Breast cancer is the most common cancer in the UK \cite{CancerReport2014UK}. The risk of being diagnosed with breast cancer is $1$ in $8$ for women in the UK and US \cite{CancerReport2014UK,CancerReport2014US}. Breast cancer is also stated to be a leading cause of cancer death in the less developed countries \cite{CancerReport2013French}.}. The case of no model mismatch is investigated first, and then the robust design scheme is applied where its power distribution in the worst-case model is evaluated and compared to the nonrobust formulation.

\emph{Notation:} Boldface (lower case) is used for column vectors, $\mathbf{x}$, and
(upper case) for matrices, $\mathbf{X}$. $ \| \a \|_{\mathbf{W}} \triangleq \sqrt{\a^H \mbf{W} \a}$ where $\mbf{W} \succ \mbf{0}$. $\mbf{x}^T$ and $\mbf{x}^H$ denote transpose and Hermitian transpose. $\mbf{R} \succeq \mbf{0}$ signifies positive semidefinite matrix and $\mbf{R}^{1/2}$ a matrix square-root, e.g., Hermitian. The set of complex numbers is denoted by $\mathcal{C}$.

\emph{Abbreviations:} Semi-infinite programming (SIP); multiple-input multiple-output (MIMO); Semidefinite program (SDP); linear matrix inequality (LMI).

\section{system model} \label{sec:system model}

We consider an array of $M$ acoustic transducers to heat target points. These transducers are located at known positions $\th_m$, for $m = 1,2,\dots,M$, around the tissue at risk, cf., \cite{08:Guo,14:Shariati2014c}. We parameterize an arbitrary point in 3D space using Cartesian coordinates $\mathbf{r} = [x \: y \: z]^T$.

Let $x_m(n)$ denote the baseband representation of narrowband discrete-time signal transmitted at the $m$th transducer, at sample $n = 1, \dots, N$. Then the baseband signal received at a generic location $\r$ equals the superposition of signals from all $M$ transducers, i.e.,
\begin{equation}\label{eq:baseband_recieved}
\begin{aligned}
y(\r,n) &= \sum_{m=1}^M  a_m(\r) x_m(n)  , \hspace{.25cm} n = 1, \ldots, N& \\
&= \a^H(\r) \x(n)  , \hspace{.25cm} n = 1, \ldots, N,&
\end{aligned}
\end{equation}
where the $m$th signal is attenuated by a factor $a_m(\mbf{r})$ which depends on the properties of the transducers, the carrier wave and the tissue. This factor is modeled as
\begin{equation}\label{eq:a_m(r)}
a_m(\r) = \frac{e^{-j2\pi f_c \tau_m(\r)}}{\|\th_m - \r\|^{\frac{1}{2}}},
\end{equation}
where $f_c$ is the carrier frequency, and $\tau_m(\r) = \frac{\|\th_m - \r\|}{c}$ is the required time for any signal to arrive at location $\r$ where $c$ is the sound speed inside the tissue. Note that the root-squared term in the denominator in \eqref{eq:a_m(r)} represents the distance dependent propagation attenuation of the acoustic waveforms.
In \eqref{eq:baseband_recieved}, the narrowband signals are represented in vector form $\mathbf{x}(n) = [x_1(n) \: \ldots \: x_m(n) \: \ldots \: x_M(n)]^T \in \mathcal{C}^{M \times 1}$ and $\a(\mbf{r}) \triangleq [a_1(\r) \: \ldots \: a_m(\r) \: \ldots a_M(\r)]^T \in \mathcal{C}^{M \times 1}$ is the array steering vector as a function of $\r$.

At a generic location $\r$  in the tissue, the power of the transmitted signal, i.e., \emph{the transmit beampattern}, is given by
\begin{equation}\label{eq:transmit_beampattren}
p(\r) = \mathbb{E} \{ | y(\r,n) |^2 \} = \a^H(\r)\R \a(\r),
\end{equation}
where
\begin{equation*}
\R \triangleq \mathbb{E} \{ \x(n) \x^H(n)\}
\end{equation*}
is the $M \times M$ covariance matrix of the signal $\x(n)$. As equation \eqref{eq:transmit_beampattren} suggests, the transmit beampattern is dependent on the waveform covariance matrix $\R$ and the array steering vector $\a(\r)$. In the following we analyze how one can form and control the  beampattern by optimizing the covariance matrix $\R$, so as to heat up the tumor region of the tissue while keeping the power deposition in the healthy tissue minimal. In this work, we consider schemes which allow for the lowest possible power leakage to the healthy area.

Once an optimal covariance matrix $\R$ has been determined, the waveform signal $\x(n)$ can be synthesized accordingly. One simple approach is $\x(n) = \R^{1/2} \mathbf{w}(n)$, where $\mathbf{w}(n)$ is a sequence of independent random vectors with mean zero and covariance matrix $\mathbf{I}$. For detailed discussion see \cite{Stoica2007b,08:Fuhrmann}\cite[ch.~14]{12:He}.

A significant challenge to this approach, however, is that the \emph{true} steering vector $\a(\r)$ in \eqref{eq:transmit_beampattren} does not exactly match the model in \eqref{eq:a_m(r)} for a host of reasons: array calibration imperfections, variations in transducing elements, tissue inhomogeneities, inaccurately specified propagation velocity, etc. We will therefore consider the aforementioned design problem subject to model uncertainties in the array steering vector at any given point $\r$. We refer to this approach as robust waveform diversity.

\section{problem formulation} \label{sec:problem formulation}

The waveform-diversity-based technique \cite{04:Fuhrmann,Stoica2007a,Stoica2007b,08:Guo,08:Fuhrmann,14:Shariati2014c} have been used for designing beampatterns \eqref{eq:transmit_beampattren} subject to practical constraints. In general, we aim to control and shape the spatial power distribution at a set of target points while simultaneously minimizing power leakage in the remaining area. By exploiting a combination of different waveforms in \eqref{eq:baseband_recieved}, the degrees of freedom increase for optimizing the beampattern under constraints.

After reviewing the standard waveform diversity approach, we focus on the practical scenario where the assumed array steering vector model is subject to perturbations. In the subsequent section, the proposed robust technique is evaluated by numerical simulations, comparing the performance with and without robustified solution under perturbed steering vectors.

\subsection{Waveform Diversity based Ultrasound System}
In the MIMO radar literature, sidelobe minimization is a beampattern design problem that has been addressed by using the waveform diversity methodology, cf., \cite{04:Fuhrmann,Stoica2007a,Stoica2007b,08:Fuhrmann}. This design problem can be thought of as an optimization problem where the probing waveforms covariance matrix $\R$ is the optimization variable to be chosen under positive semi-definiteness assumption and with a constraint on the total power. The waveform-diversity-based scheme for ultrasound system has been introduced and explained in detail in \cite{08:Guo} based on the transmit beampattern design technique for MIMO radar systems \cite{04:Fuhrmann,Stoica2007a}.

In the following we consider the practical power constraint where all array elements have the same power. Therefore, the covariance matrix $\R$ belongs to the following set $\mathcal{R}$:
\begin{equation}
\mathcal{R} \triangleq \{\mathbf{R}
\hspace{.1cm}|\hspace{.1cm} \mathbf{R}\succeq \mbf{0} , R_{mm} =
\frac{\gamma}{M}, m=1,2,...,M \},
\end{equation}
where $\gamma$ is the total transmitted power and $R_{mm}$ is the $m$th diagonal element of $\R$ corresponding to the power emitted by $m$th transducer. The healthy tissue and the tumor regions are represented by two sets of discrete control points $\r$:
\begin{equation*}
\begin{split}
\Omega_S &= \{ \r_1,\r_2,\ldots,\r_{N_S}\} \\
\Omega_T &= \{ \r_1,\r_2,\ldots,\r_{N_T} \},
\end{split}
\end{equation*}
where $N_S$ and $N_T$ denote the number of points in the healthy tissue region and the tumor regions, respectively. Without loss of generality, let $\r_0$ be a representative point which is taken to be the center of the tumor region $\Omega_T$.  The objectives for this optimization problem can be summarized as follows: Design the waveform covariance matrix $\mbf{R}$ so as to
\begin{itemize}
\item maximize the gap between the power at the tumor center $\r_0$ and the power at the control points $\r$ in the healthy tissue region $\Omega_S$;
\item while guaranteeing a certain power level for control points $\r$ in the tumor region $\Omega_T$.
\end{itemize}

Mathematically, this problem is formulated as (see \cite{08:Guo})
\begin{equation}\label{eq:sidelobe min}
\begin{aligned}
\underset{\R,t}{\textrm{max}}  \hspace{.5cm} &t& \\
\textrm{s.t.}  \hspace{.5cm} &\a^H(\r_0) \R \a(\r_0) -  \a^H(\r) \R \a(\r) \geq t, \forall \r \in \Omega_S &\\
& \a^H(\r) \R \a(\r)  \geq (1 - \delta) \a^H(\r_0) \R \a(\r_0), \forall \r \in \Omega_T& \\
& \a^H(\r) \R \a(\r) \leq  (1 + \delta) \a^H(\r_0) \R \a(\r_0), \forall \r \in \Omega_T&\\
&\R \in \mathcal{R}&
\end{aligned}
\end{equation}
where $t$ denotes the gap between the power at $\r_0$ and the power at the control points $\r$ in the healthy region $\Omega_S$. The parameter $\delta$ is introduced here to control the required certain power level at the control points in the tumor region. For instance, if we set $\delta = 0.1$, then we aim for having power at the tumor region $\Omega_T$ to be within 10\% of $p(\r_0)$, i.e., the power at the tumor center. This is an SDP problem which can be solved efficiently in polynomial time using any SDP solver, e.g., \texttt{CVX} \cite{cvx2013,gb08}.

\subsection{Robust Waveform Diversity based Ultrasound System}

The convex optimization problem \eqref{eq:sidelobe min} and consequently its optimal solution, i.e., the optimal covariance matrix $\mbf{R}$, are functions of the steering vectors $\a(\r)$. In practice, however, the assumed steering vector model used to optimize $\mbf{R}$ is inaccurate. Hence using \emph{nominal} steering vectors $\ahat(\r)$ based on an ideal model, in lieu of the unknown \emph{true} steering vectors $\a(\r)$ in \eqref{eq:sidelobe min}, may result in undesired beampatterns with low power at the tumor region and damaging power deposition in the healthy tissue region. Such health considerations in medical applications motivate an approach that is robust with respect to the worst-case model uncertainties.

In order to formulate the robust design problem mathematically, we parameterize the steering vector uncertainties as follows. Let the true steering vector for the transducer array be $\mathbf{a(\r)} = \mathbf{\hat{a}(\r)} + \mathbf{\tilde{a}(\r)}$ where  $\mathbf{\tilde{a}(\r)}$ is an unknown perturbation from the nominal steering vector. The deterministic perturbation at any generic point $\r$ belongs to uncertainty set $\mathcal{E}_{\r}$ that is bounded
\begin{equation*}
\begin{split}
\mathcal{E}_{\r} \triangleq \{ \mathbf{\tilde{a}}(\r) \hspace{.2cm} | \hspace{.2cm} \| \mathbf{\tilde{a}}(\r) \|_{\mathbf{W}}^2 \leq \epsilon_{\r} \},
\end{split}
\end{equation*}
where $\mathbf{W}$ is an $M \times M$ diagonal weight matrix with positive elements. The weight matrix $\mathbf{W}$ can be derived based on the type of uncertainty. Using $\mathbf{W}$, the set $\mathcal{E}_{\r}$ indicates an ellipsoidal region. The bound $\epsilon_{\r}$ for the set can be a constant or a function of $\r$, i.e., $\epsilon_{\r} = f(\r)$. This set enables parameterization of element-wise uncertainties in the nominal steering vector $\mathbf{\hat{a}(\r)}$ at each $\r$.

Besides this consideration, we generalize the problem formulation \eqref{eq:sidelobe min} further by setting a uniform bound (power level) $P$ across the tumor region $\Omega_T$ as an optimization variable to which the power of all the control points in the healthy region $\Omega_S$ are compared. This is in contrast to \eqref{eq:sidelobe min} and the robust formulation in \cite{14:Shariati2014c}, where the power levels of all the healthy grid points $\Omega_S$ are compared with the power of only a single reference point at fictitious tumor center $\r_0$. There is no need to limit our problem to a single point as a reference power level. Rather, the desired tightness of the power level across $\Omega_T$ is specified by the parameter $0 \leq \delta < 1$. This generalization also improves the efficiency when it comes to solving the robust design problem. 

With these considerations, the robust beampattern design problem can be formulated as\begin{equation}\label{eq:robust beampattern problem}
 \begin{aligned}
\underset{\R,t,P}{\textrm{max}} \hspace{.3cm} &t \hspace{.3cm} \textrm{subject to}&\\
&\hspace{-1cm} P - \left(\ahat(\r) \!\!+\!\! \atil(\r)\right)^H  \R  \left(\ahat(\r) \!\!+\!\! \atil(\r)\right) \geq t, \forall \atil(\r) \in \mathcal{E}_{\r}, \r \in \Omega_S&\\
&\hspace{-1.3cm}\left(\ahat(\r) \!\!+\!\! \atil(\r)\right)^H  \R  \left(\ahat(\r) \!\!+\!\! \atil(\r)\right) \geq (1-\delta)P, \forall \atil(\r) \in \mathcal{E}_{\r}, \r \in \Omega_T&\\
&\hspace{-1.3cm}\left(\ahat(\r) \!\!+\!\! \atil(\r)\right)^H  \R  \left(\ahat(\r) \!\!+\!\! \atil(\r)\right) \leq (1+\delta)P, \forall \atil(\r) \in \mathcal{E}_{\r}, \r \in \Omega_T&\\
&\hspace{-1cm}\R \in \mathcal{R},&
\end{aligned}
\end{equation}
where $t$ is the gap between the desired power level set across $\Omega_T$ and power deposition in the healthy tissue $\Omega_S$, similar to \eqref{eq:sidelobe min}.
Note that we take into account every possible perturbation $\atil(\r) \in \mathcal{E}_{\r}$.

In contrast to the optimization problem \eqref{eq:sidelobe min} which is a tractable convex problem, the robust problem \eqref{eq:robust beampattern problem} is an SIP problem. For a given $\R$ in \eqref{eq:robust beampattern problem}, there are infinite number of constraints in terms of $\atil(\r)$ to satisfy which makes the problem non-trivial. However, in the following theorem, extending the approach in \cite{14:Shariati2014c}, we reformulate the robust power deposition problem as a convex SDP problem whose solution is the optimally robust covariance matrix.
\begin{theo}\label{theo:robustSDP}
The robust power deposition for an M-element transducer array with the probing signal covariance matrix $\R \in \mathcal{R}$ and the perturbation vector $\atil(\r) \in \mathcal{E}_{\r}$, i.e., the solution of \eqref{eq:robust beampattern problem}, is given as a solution to the following SDP problem
\begin{equation}\label{eq:robustSDP}
\begin{aligned}
&\underset{\R,t,P,\beta_i,\beta_{j,1},\beta_{j,2}}{\textrm{max}} \hspace{.3cm} t \hspace{.5cm} \textrm{subject to}& \\
&\Omega_S\!\!:\!\!\left[ \begin{array}{cc}
\beta_i \mathbf{W} - \R   &    -\R \ahat(\r_i) \\
-\ahat(\r_i)^H \R         &     P - t - \ahat(\r_i)^H \R \ahat(\r_i) - \beta_i \epsilon_{\r_i}
\end{array} \right] \succeq \mbf{0}, &\\
&\Omega_T\!\!:\!\!\left[ \begin{array}{cc}
\beta_{j,1} \mathbf{W} + \R   &    \R \ahat(\r_j) \\
\ahat(\r_j)^H \R              &    \ahat(\r_j)^H \R \ahat(\r_j) \!\!-\!\! (1-\delta)P \!\!-\!\! \beta_{j,1} \epsilon_{\r_j}
\end{array} \right] \succeq \mbf{0}, &\\
&\Omega_T\!\!:\!\!\left[ \begin{array}{cc}
\beta_{j,2} \mathbf{W} - \R   &    -\R \ahat(\r_j) \\
-\ahat(\r_j)^H \R             &     (1+\delta)P \!\!-\!\! \ahat(\r_j)^H \R \ahat(\r_j) \!\!-\!\! \beta_{j,2} \epsilon_{\r_j}
\end{array} \right] \succeq \mbf{0}, &\\
&\R \in \mathcal{R},\beta_i,\beta_{j,1},\beta_{j,2} \geq 0,  i=1,\ldots,N_S, j=1,\ldots,N_T.&
\end{aligned}
\end{equation}

\end{theo}

\emph{Proof:} See Appendix~\ref{app A}.

Observe that the notations $\Omega_S$ and $\Omega_T$ indicate that the corresponding linear matrix inequalities (LMIs) should be satisfied for the points $\r_i \in \Omega_S$ and $\r_j \in \Omega_T$, respectively. Note that the robust SDP problem in this paper, which is stated in Theorem~\ref{theo:robustSDP}, can be solved more efficiently than the SDP problem in \cite{14:Shariati2014c} since the matrices $\R$ and $\mathbf{W}$ in the current formulation have half of the size of the matrices involved in the latter problem. This occurs due to the generalization of the robust problem by using the uniform power level as a benchmark.

Note that other robust problems with similar objectives can also be addressed using the above approach which are outlined in the following subsection. 

\subsection{Alternative robust formulations} \label{subsec:alternative}

Similar robust problems to that of \eqref{eq:robust beampattern problem} can be formulated in many different ways. For example, by restricting the power level outside the tumor in a weighted fashion.
\begin{eqnarray*}
\min_{t,R}&& t \hspace{.3cm} \mbox{subject to}\\
&& \hspace{-1.3cm}(\ahat(\r)\!\!+\!\!\atil (\r))^H R (\ahat(\r) \!\!+\!\! \atil (\r)) \le t w(\r), \forall \atil (\r)\in \mathcal{E}_\r, \r\in \Omega_{\rm S}\\
&&\hspace{-1.3cm} (\ahat (\r)\!\!+\!\!\atil (\r))^H R (\ahat(\r) \!\!+\!\! \atil (\r)) \ge (1-\delta)P, \forall \atil (\r)\in \mathcal{E}_\r, \r\in \Omega_{\rm T}\\
&& \hspace{-1.3cm}(\ahat (\r)\!\!+\!\!\atil (\r))^H R (\ahat(\r) \!\!+\!\! \atil (\r)) \le (1+\delta)P, \forall \atil (\r)\in \mathcal{E}_\r, \r\in \Omega_{\rm T}\\
&&\hspace{-1.3cm}R\in \mathcal{R}
\end{eqnarray*}
where $P,\delta$ are fixed and $w(\r)$ is a weighting function constructed, e.g., so that the energy bound close to the tumor is less restrictive.

One could also construct problems that minimize the sum of the energy in the non-tumor area where $t(\r)$ denotes the energy at $\r$:
\begin{eqnarray*}
\min_{t(\r),R}&& \sum_{\r\in\Omega_{\rm S}}t(\r)  \hspace{.3cm} \mbox{subject to} \\
&& \hspace{-1.3cm} (\ahat(\r) \!\!+\!\! \atil (\r))^H R (\ahat(\r) \!\!+\!\! \atil (\r)) \le t(\r), \forall \atil (\r)\in \mathcal{E}_\r, \r\in \Omega_{\rm S}\\
&& \hspace{-1.3cm} (\ahat(\r) \!\!+\!\! \atil (\r))^H R (\ahat(\r) \!\!+\!\! \atil (\r)) \ge (1-\delta)P, \forall \atil (\r)\in \mathcal{E}_\r, \r\in \Omega_{\rm T}\\
&& \hspace{-1.3cm} (\ahat(\r) \!\!+\!\! \atil (\r))^H R (\ahat(\r) \!\!+\!\! \atil (\r)) \le (1+\delta)P, \forall \atil (\r)\in \mathcal{E}_\r, \r\in \Omega_{\rm T}\\
&&\hspace{-1.3cm} R\in \mathcal{R}.
\end{eqnarray*}
Both of the alternative formulations described above can be addressed following the steps derived in Appendix~\ref{app A} by using $\mathcal{S}$-lemma since we are still dealing with quadratic constraints.

In the next section, we illustrate the reference performance of a nominal scenario where the steering vectors are perfectly known. Then we observe how much power can leak to the healthy tissue and cause damages when subject to uncertain steering vectors. Finally, we evaluate the proposed robust scheme in terms of improving the power deposition along our stated design goals.

\section{Numerical Results} \label{sec:numerical_results}

\begin{figure}
   \begin{center}
   \includegraphics[width=\columnwidth,height=5.5cm]{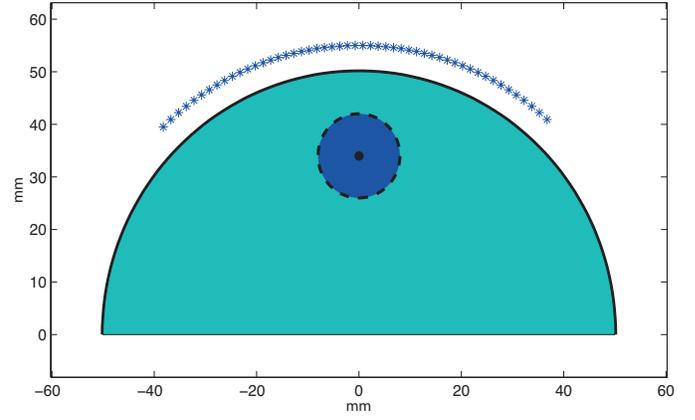}\\ 
 \caption{A schematic 2-D breast model with an $16$-mm embedded tumor at $(0,34)$ as a reference geometry. A curvilinear ultrasonic array with $51$ transducers is located near to the organ at risk. The ultrasonic array is used for hyperthermia treatment.}
   \label{fig:2D_model}
   \end{center}
\end{figure}

To illustrate the performance of the proposed robust scheme, we consider a 2D model of the organ at risk. Here, similar to \cite{08:Guo}, we focus on the ultrasonic hyperthermia treatment for breast cancer where a 10-cm-diameter semi-circle is assumed to model breast tissues with a 16-mm-diameter tumor embedded inside. The tumor center is located at $\r_0 = [0 \hspace{.1cm} 34]^T$ mm. Fig.~\ref{fig:2D_model} shows this schematic model. We consider a curvilinear array with $M=51$ acoustic transducers and half wavelength element spacing. Acoustic waveforms used to excite the array have the carrier frequency of $500$ kHz. The acoustic wave speed for the breast tissue is considered $1500$ m/s.

To characterize (discretize) the healthy tissue region $\Omega_S$ and the tumor region $\Omega_T$, two grid sets with the spacing $4$mm are considered. For optimization, a rectangular surface of the dimension $ 64 \times 42$ in mm is assumed symmetric around the tumor to model the healthy region $\Omega_S$, while the grid points belonging to the circular tumor region are excluded from this surface and they model $\Omega_T$. Overall, $174$ and $13$ number of control points are considered to characterize $\Omega_S$ and $\Omega_T$ in order to optimize the array beampattern. 

The total transmitted power is constrained to $\gamma = 1$. For simplicity, the uncertainty set $\mathcal{E}_{\r}$ is modeled with $\mbf{W} = \mbf{I}_M$ with $\epsilon_{\r} \equiv \epsilon$ for all $\r$ where $\epsilon = 0.25$. Furthermore, the tightness of the desired power level in the across tumor region, $\delta$, is set to $0.7$. Note that for the small values of $\delta$ and/or large values of $\epsilon$, the problem may turn infeasible. In general, the feasibility of the problem depends on the value of the tightness bound $\delta$ relative to the size of the existing uncertainty in the system, i.e., the volume of the uncertainty set $\epsilon$, and the number of grid points $N_S$ and $N_T$ used to control the beampattern at the area of interest. When $\delta$ is too small, the desired power level across $\Omega_T$ is close to uniform and there may not exist enough degrees of freedom for the design problem to have a solution. 

For reference, the optimal covariance matrix when no uncertainty is taken into account, $\mathbf{R}_{nr}$, is obtained by solving problem \eqref{eq:robust beampattern problem} using only nominal steering vectors $\ahat(\r)$, i.e.,  $\atil(\r) \equiv \mbf{0}$. The optimal robust covariance matrix, denoted $\R^\star$, is obtained by solving \eqref{eq:robustSDP} where $\atil(\r) \in \mathcal{E}_{\r}$. For performance evaluation, we consider the power deposition in the tissue under the worst-case perturbations of the steering vectors. This scenario provides a lower bound to the achievable performances of all steering vector perturbations $\atil(\r)$ which belong to the deterministic uncertainty set $\mathcal{E}_{\r}$. In other words, for the points $\r$ in the healthy region $\Omega_S$, the worst-case performance is rendered by the steering vectors which provide the highest power, whereas for the points $\r$ in the tumor region $\Omega_T$, those steering vectors which attain the lowest power are the ones which contribute in the worst-case performance. They are collectively referred to as the \textit{worst steering vectors}. Therefore, for a given $\R$, either $\R_{nr}$ or $\R^\star$, the worst steering vectors for the control points $\r$ in $\Omega_S$ and $\Omega_T$, are obtained by maximizing and minimizing the transmit beampattern \eqref{eq:transmit_beampattren}, respectively. Observe that finding the worst steering vectors for the points in the tumor region $\Omega_T$ equals solving the following convex minimization problem at each $\r \in \Omega_T$, i.e.,
\begin{equation*}
\underset{\|\atil(\r)\|^2 \leq \epsilon}{\textrm{min}} \hspace{.2cm} (\ahat(\r) + \atil(\r))^H \R (\ahat(\r) + \atil(\r))
\end{equation*}
using \texttt{CVX} \cite{cvx2013,gb08}.
Whereas, for finding the worst steering vectors for the points in the healthy region $\Omega_S$, we obtain a local optimum for the following non-convex maximization problem at each $\r \in \Omega_S$, i.e.,
\begin{equation*}
\underset{\|\atil(\r)\|^2 \leq \epsilon}{\textrm{max}} \hspace{.2cm} (\ahat(\r) + \atil(\r))^H \R (\ahat(\r) + \atil(\r)),
\end{equation*}
using semidefinite relaxation techniques from \cite{Beck06strongduality}.

\begin{figure}
   \begin{center}
   \includegraphics[width=\columnwidth,height=5.5cm]{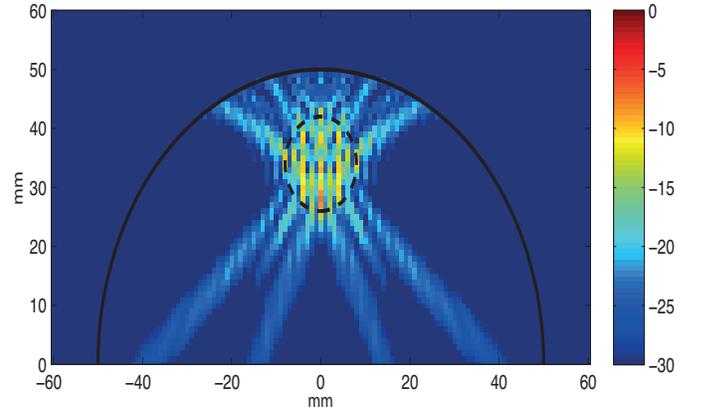}\\ 
 \caption{Power distribution (transmit beampattern in dB) for the nominal scenario, i.e., using $\R_{nr}$ and $\atil(\r) \equiv 0$.}
   \label{fig:nominal}
   \end{center}
\end{figure}

We evaluate the designed beampatterns \eqref{eq:transmit_beampattren} plotting the spatial power distribution in decibel scale, i.e., $20 \log_{10} (p(\r))$. Two different scenarios are considered, namely, \textit{nominal} and \textit{perturbed}, to evaluate the proposed robust power distribution scheme for the ultrasonic array. In the first scenario, nominal, we assume that the array steering vectors are precisely modeled, i.e., $\atil(\r) = \mathbf{0}$. In Fig.~\ref{fig:nominal}, the beampattern generated by the array is plotted for the nominal scenario. This figure represents how power is spatially distributed over the organ at risk in an idealistic situation. Here, the covariance matrix of the waveforms is optimized under the assumption that the steering vectors are accurately modeled by \eqref{eq:a_m(r)}, and the performance is evaluated using exactly the same steering vectors without any perturbations. The power is noticeably concentrated in the tumor region and importantly the power in the healthy tissue is several decibels lower.

\begin{figure}
   \begin{center}
   \includegraphics[width=\columnwidth,height=5.5cm]{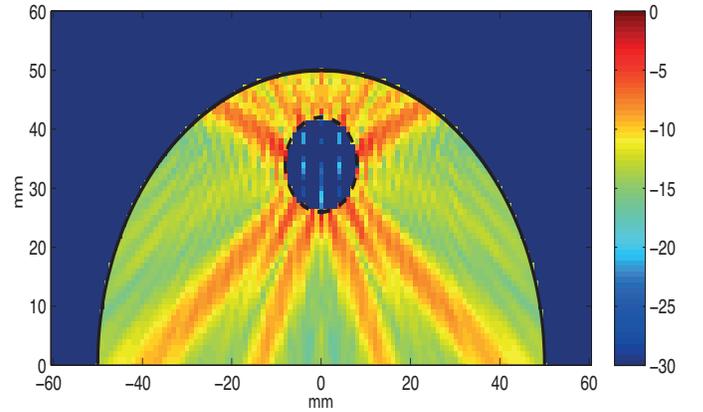}\\ 
 \caption{Power distribution (transmit beampattern in dB) for the perturbed scenario, i.e., using $\R_{nr}$ and $\atil(\r) \in \mathcal{E}_{\r}$.}
   \label{fig:NonRobustWorst}
   \end{center}
\end{figure}

\begin{figure}
   \begin{center}
   \includegraphics[width=\columnwidth,height=5.5cm]{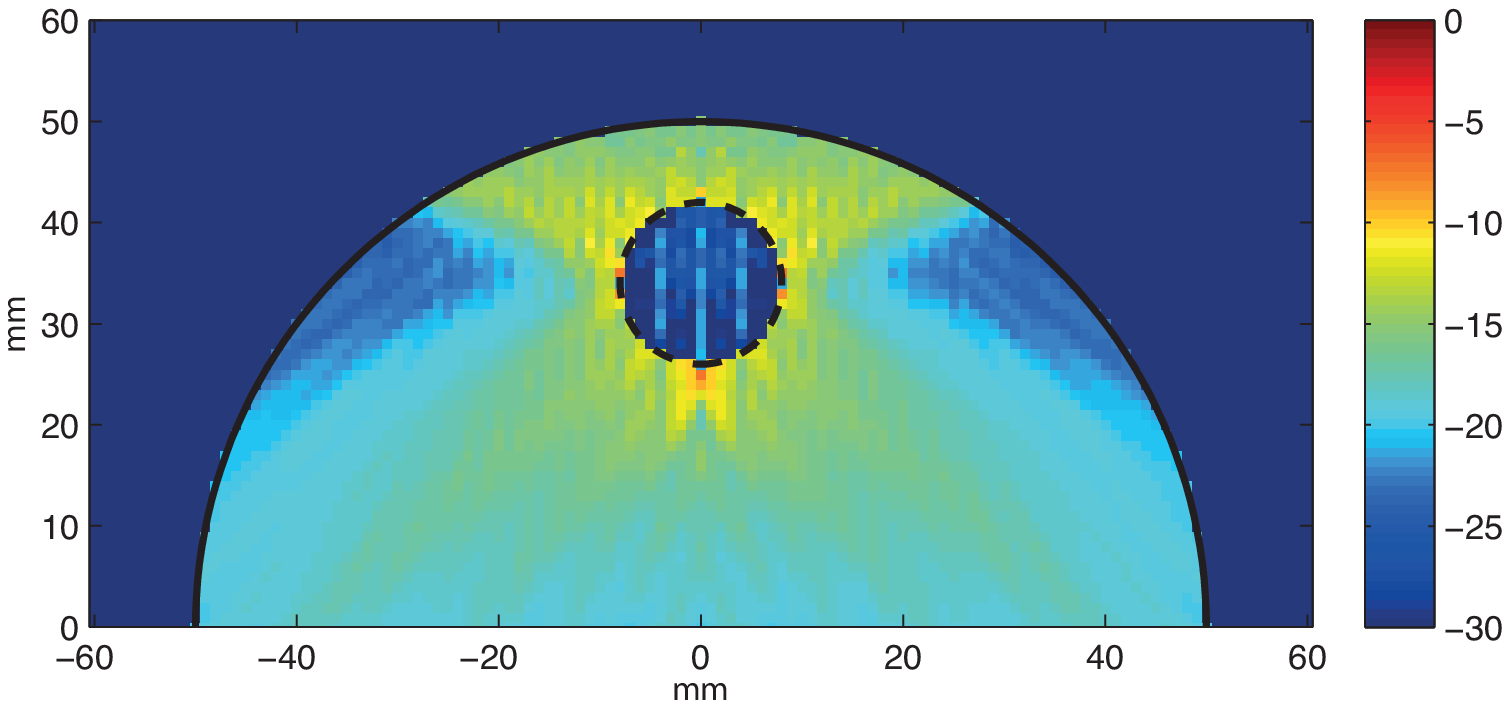}\\
 \caption{Power distribution (transmit beampattern in dB) for the perturbed scenario, i.e., using $\R^\star$ and $\atil(\r) \in \mathcal{E}_{\r}$.}
   \label{fig:RobustWorst}
   \end{center}
\end{figure}

In the second scenario, perturbed, the idealistic assumptions are relaxed and model uncertainties and imperfections are taken into account. The second scenario represents the case where the true steering vectors are perturbed versions of the nominal steering vectors $\ahat(\r)$, i.e., the true steering vector equals $\ahat(\r) + \atil(\r)$ where $\atil(\r) \in \mathcal{E}_{\r}$. The perturbation vectors $\atil(\r)$ are unknown but deterministically bounded. In the following we illustrate the worst-case performance, i.e., using the worst steering vectors to calculate the power distribution at each point. We start by illustrating the beampattern for the non-robust covariance matrix $\R_{nr}$ under the worst steering vectors. Fig.~\ref{fig:NonRobustWorst} shows how steering vector errors can degrade the array performance. Notice that in the worst-case, there is a substantial power leakage that occurs in the healthy tissue surrounding the tumor compared to Fig.~\ref{fig:nominal}. While, in Fig.~\ref{fig:RobustWorst}, the robust optimal covariance matrix $\R^\star$, i.e., the solution to \eqref{eq:robustSDP}, is used to calculate the power for the worst steering vectors. Comparing Fig.~\ref{fig:NonRobustWorst} and Fig.~\ref{fig:RobustWorst}, we see that by taking model uncertainties into account it is possible to obtain a noticeable increase in power in the tumor region for the worst case, and importantly, dramatic reductions of power deposited in the healthy tissue.

To finalize the numerical analysis, we provide a quantitative description for the performance of our proposed scheme summarized in Table~\ref{table:power}. It shows the average power calculated in dB received at the tumor region $\Omega_T$ and at the healthy region $\Omega_S$.

\vspace{-0.4cm}
\begin{table}[h!]
\caption{Average power for different regions}
     \label{table:power}    \vspace{-0.4cm}
\begin{center}
     \begin{tabular}{ | c | c | c |}
     \hline
     Scenarios &  $\Omega_T$ & $\Omega_S$   \\ \hline \hline
     Nominal, $\R_{0}$ & $-16.54$ & $-29.78$  \\ \hline
     Perturbed, $\R_{0}$ & $-36.40$ & $-11.69$  \\ \hline
     Perturbed, $\R^\star$ & $-27.17$ & $-17.43$  \\ \hline
     \end{tabular}
         \end{center}
           \vspace{-0.4cm}
\end{table}

\section{Conclusion}

The robust transmit signal design for optimizing spatial power distribution of an multi-antenna array is investigated. A robustness analysis is carried out to combat against inevitable uncertainty in model parameters which results in performance degradation. Such degradation occurs in practice quite often due to relying on imperfect prior and designs based upon them. Particularly, in this paper, the transmit signal design is based on exploiting the waveform diversity property but where errors in the array steering vector are taken into account. These errors are modeled as belonging to a deterministic set defined by a weighted norm. Then, the resulting robust signal covariance optimization problem with infinite number of constraints is translated to a convex problem which can be solved efficiently, by using the $\mathcal{S}$-procedure.

Designs that are robust with respect to the worst case are particularly vital in biomedical applications due to health risks and possible damage. Herein we have focused on local hyperthermia therapy as one of the cancer treatments to be used either individually or along with other treatments such as radio/chemotherapy. Specifically, we consider hyperthermia treatment of breast cancer motivated by the fact that breast cancer is a major global health concern. The proposed robust signal design scheme aims to reduce unwanted power leakage into the healthy tissue surrounding the tumor while guaranteeing certain power level in the tumor region itself.

We should emphasize on the fact that the robust design problem formulation and the analysis carried out herein yielding to the robust waveforms are general enough to be exploited whenever spatial power distribution is a concern to be addressed in real world scenarios dealing with uncertainties, e.g., for radar applications.

Numerical examples representing different scenarios are given to illustrate the performance of the proposed scheme for hyperthermia therapy. We have observed significant power leakage into the healthy tissue that can occur if the design is based on uncertain model parameters. Importantly, we have shown how such damaging power deposition can be avoided using the proposed robust design for optimal spatial power distribution.

\section{Acknowledgement}
The authors would like to acknowledge Prof. Jian Li for providing an implementation of examples from \cite{08:Guo}.

\appendix \label{sec:appendix}

\subsection{Proof of Theorem~\ref{theo:robustSDP}} \label{app A}

We start the proof by first stating the $\mathcal{S}$-Procedure lemma which helps us to turn the optimization problem \eqref{eq:robust beampattern problem} with infinitely many quadratic constraints into a convex problem with finite number of LMIs.

\begin{lem}\label{lem 6}($\mathcal{S}$-Procedure \cite[Lemma 4.1]{Beck2009a}):
Let $f_k(\x): \mathbb{C}^n \rightarrow \mathbb{R}$, $k = 0,1$, be defined as
$f_k(\x) = \x^H\mathbf{A}_k \x
+ 2 \textrm{Re} \{\mathbf{b}_k^H \x \} + c_k$, where $\mathbf{A}_k = \mathbf{A}_k^H \in \mathbb{C}^{n
\times n}, \mathbf{b}_k \in \mathbb{C}^n$, and $c_k \in
\mathbb{R}$. Then, the statement (implication) $f_0(\x) \geq 0$ for all $\x \in \mathbb{C}^n$ such that $f_1(\x) \geq 0$ holds if and only if there
exists $\beta \geq 0$ such that\footnote{Note that $\mathcal{S}$-Procedure is lossless in complex space for the case of at most two constraints \cite{01:Jonsson}.}
\begin{equation*}
\left[ \begin{array}{cc}
\mathbf{A}_0 & \mathbf{b}_0 \\
\mathbf{b}_0^H & c_0
\end{array}\right] - \beta \left[ \begin{array}{cc}
\mathbf{A}_1 & \mathbf{b}_1 \\
\mathbf{b}_1^H & c_1
\end{array}\right] \succeq 0,
\end{equation*}
if there exists a point $\mathbf{\hat{x}}$ with
$f_1(\mathbf{\hat{x}}) > 0$.
\end{lem}

The constraints in the optimization problem \eqref{eq:robust beampattern problem} can be rewritten as the following functions of $\atil(\r)$ for $\r \in \Omega_S$ and $\r \in \Omega_T$. For notation simplicity we only specify the set from which the control points are drawn, and we also drop $\r$.
\[
\Omega_S:
\begin{cases}
 f_0 = -\atil^H \R \atil - 2 \textrm{Re}(\ahat^H \R \atil) - \ahat^H \R \ahat - t + P \geq 0 \\
 f_1 = - \atil^H \mathbf{W} \atil + \epsilon_{\r} \geq 0
\end{cases}
\]
\[
\Omega_T:
\begin{cases}
 f_0 = \atil^H \R \atil + 2 \textrm{Re}(\ahat^H \R \atil) + \ahat^H \R \ahat - (1-\delta)P \geq 0 \\
 f_1 = - \atil^H \mathbf{W} \atil + \epsilon_{\r} \geq 0
\end{cases}
\]
\[
\Omega_T:
\begin{cases}
 f_0 = -\atil^H \R \atil - 2 \textrm{Re}(\ahat^H \R \atil) - \ahat^H \R \ahat  + (1+\delta)P \geq 0 \\
 f_1 = - \atil^H \mathbf{W} \atil + \epsilon_{\r} \geq 0
\end{cases}
\]
Now, according to the $\mathcal{S}$-Procedure lemma, each pair of the quadratic constraints above is replaced with an LMI for each grid points in the pre-defined sets. In other words, all these quadratic constraints are satisfied simultaneously if we find $\beta_i$ for $i=1,\ldots,N_S$, $\beta_{j,1}$ and $\beta_{j,2}$ for $j=1,\ldots,N_T$ for which the mentioned LMIs in Theorem~\ref{theo:robustSDP} holds. Thus, the problem boils down to the SDP problem \eqref{eq:robustSDP} with $2N_T + N_S$ LMIs of the size $(M+1) \times (M+1)$ as the constraints. \hspace{2cm}$\Box$

\bibliographystyle{IEEEtran}
\bibliography{bibliokthNafis}

\end{document}